\documentclass[aps,prl,twocolumn,showpacs,superscriptaddress]{revtex4-1} 
\usepackage[utf8]{inputenc}
\usepackage[english]{babel}
\usepackage{graphicx}  
\usepackage{dcolumn}   
\usepackage{bm}        
\usepackage{amssymb,amsmath}   
\usepackage{hyperref}
\usepackage{cleveref}
\usepackage{color}
\usepackage[dvipsnames]{xcolor}

\newcommand{\h}{\left(\frac12\right)^{\frac13}}
\newcommand{\tT}{\tilde{T}}

\newcommand{\tTin}{\tilde{T}^{\rm{in}}}
\newcommand{\tTex}{\tilde{T}^{\rm{ex}}}
\newcommand{\ttt}{\tilde{t}}
\newcommand{\tC}{\tilde{C}}
\newcommand{\tin}{t_{\rm{in}}}
\newcommand{\Cin}{{C}^{\rm{in}}}
\newcommand{\Cex}{{C}^{\rm{ex}}}
\newcommand{\Cmu}{{C}^{\rm{\mu}}}
\newcommand{\tCin}{\tilde{C}^{\rm{in}}}
\newcommand{\tCex}{\tilde{C}^{\rm{ex}}}
\newcommand{\tCmu}{\tilde{C}^{\rm{\mu}}}
\newcommand{\Csat}{C_{\rm sat}}
\newcommand{\Pex}{\mathcal{P}^{\rm{ex}}}
\newcommand{\Pin}{\mathcal{P}^{\rm{in}}}
\renewcommand{\P}{\mathcal{P}}
\newcommand{\tPex}{\tilde{\mathcal{P}}^{\rm{ex}}}
\newcommand{\tPin}{\tilde{\mathcal{P}}^{\rm{in}}}
\newcommand{\tP}{\tilde{\mathcal{P}}}
\newcommand{\tp}{\tilde{p}}
\renewcommand{\L}{\mathcal{L}}

\newcommand{\G}{\mathcal{G}}

\newcommand{\cb}{\color{black}}
\newcommand{\vo}{\dot{V}_{O2\text{max}}}
\newcommand{\vv}{\dot{V}_{O2}}
\newcommand{\ve}{\dot{V}_E}
\newcommand{\D}{\mathcal{D}}
\newcommand{\Do}{D\scriptstyle L \textstyle_{O_2}}
\newcommand{\Dh}{D\scriptstyle L \textstyle_{H_2O}}
\renewcommand{\DH}{D\scriptstyle L \textstyle_{\P}}
\newcommand{\ja}{j_{H_2O}}
\newcommand{\vh}{v\scriptstyle L \textstyle_{H_2O}}
\newcommand{\timeh}{t\scriptstyle L \textstyle_{H_2O}}
\newcommand{\Vh}{\dot{V}_{H_2O}}
\newcommand{\psat}{p_{\rm sat}}

\begin{document}


\title{Allometric scaling of heat and water exchanges in the mammals' lung.}

\author{B. Sobac} \affiliation{TIPs Lab (Transfers, Interfaces and Processes), Université libre de Bruxelles, Brussels, Belgium}
\author{C. Karamaoun}\affiliation{TIPs Lab (Transfers, Interfaces and Processes), Université libre de Bruxelles, Brussels, Belgium}\affiliation{Université C\^{o}te d'Azur, CNRS, LJAD, Centre VADER, Nice, France}
\author{B. Haut}\affiliation{TIPs Lab (Transfers, Interfaces and Processes), Université libre de Bruxelles, Brussels, Belgium}
\author{B. Mauroy}\affiliation{Université C\^{o}te d'Azur, CNRS, LJAD, Centre VADER, Nice, France}

\date{\today}

\begin{abstract}
Mammals have a high metabolism that produces heat proportionally to the power $3/4$ of their mass at rest. 
Any excess of heat has to be dissipated in the surrounding environment to prevent overheating.
Most of that dissipation occurs through the skin, but the efficiency of that mechanism decreases with the animal's mass. 
The role of the other mechanisms for dissipating heat is then raised, more particularly the one linked to the lung that forms a much larger surface area than the skin. 
The dissipation occurring in the lung is however often neglected, even though there exists no real knowledge of its dynamics, hidden by the complexity of the organ's geometry and of the physics of the exchanges.
Here we show, based on an original and analytical model of the exchanges in the lung, that all mammals, independently of their mass, dissipate through their lung the same proportion of the heat they produced, about $6-7\%$.
We found that the heat dissipation in mammals’ lung is driven by a number, universal among mammals, that arises from the dynamics of the temperature of the bronchial mucosa. We propose a scenario to explain how evolution might have tuned the lung for heat exchanges.  
Furthermore, our analysis allows to define the pulmonary heat and water diffusive capacities. We show in the human case that these capacities follow closely the oxygen consumption.
Our work lays the foundations for more detailed analysis of the heat exchanges occurring in the lung. Future studies should focus on refining our understanding of the universal number identified. In an ecological framework, our analysis paves the way to a better understanding of the mammals' strategies for thermoregulation and of the effect of warming environments on mammals' metabolism.

\end{abstract}

\maketitle

In mammals, including humans, feeling hot starts at temperatures that are lower than that of their body, actually somewhere about ten degrees Celsius lower~\cite{kirch_temperature_2005}.
This feeling is related to the way mammals' metabolism deals with energy.
Respiration stores our energy by metabolising adenosine triphosphate (ATP) from the oxidation of substrates~\cite{vusse_substrate_1995}. 
Only about forty percents of the energy freed by the oxidative processes are actually stored, and the sixty percents left are released as heat~\cite{weibel_pathway_1984}. When energy is needed, ATP is degraded, and, again, some of the energy produced is released as heat, with about as much as eighty percents for mammals' muscles~\cite{smith_efficiency_2005}.
The resulting amount of heat can be excessive, and thermoregulation has to eliminate the surplus~\cite{girardier_mammalian_2012} to prevent overheating, potentially deadly~\cite{yan_pathophysiological_2006, angilletta_jr_thermal_2009}. 
Numerous specific evolutionary strategies have actually emerged amongst mammals to improve heat dissipation, either behavioral~\cite{terrien_behavioral_2011} or morphological~\cite{bligh_temperature_1973, young_evidence_1982}, supporting the fact that heat dissipation might be an active evolutionary drive, as suggested by Heat Dissipation Limit theory~\cite{speakman_maximal_2010, speakman_heat_2010}.

The main mechanism used by mammals to eliminate the excessive heat from their body is based on exchanges through their skin~\cite{ingram_man_2011, kirch_temperature_2005}. This mechanism relies on the temperature and water concentration differences between their skin and the surrounding environment. This process is however not equally efficient for all mammals~\cite{speakman_maximal_2010}, as the proportion of heat dissipated relatively to the heat produced decreases with the animal's mass. Indeed, both the amount of heat dissipated through the skin and the amount of heat produced by the metabolism are following allometric scaling laws~\cite{peters_ecological_1986}, i.e. vary as a power laws of the animal mass. But the dissipation through the skin has an exponent of about $0.63$~\cite{speakman_maximal_2010}, smaller than the one of heat production, which is in the range $0.69-0.76$ at rest metabolic rate~\cite{peters_ecological_1986, west_general_1997, white_mammalian_2003, kozlowski_is_2004, brown_yes_2005, glazier_beyond_2005, roberts_new_2010, maino_reconciling_2014}, in the range $0.73-0.75$ for field metabolic rate~\cite{nagy_energetics_1999, nagy_field_2005, white_allometric_2005} and about $0.87$ at maximum metabolic rate ($\vo$)~\cite{weibel_exercise-induced_2005}. Hence, large mammals might have less amplitude for increasing their metabolic rates than small ones and have to rely on lower maximal metabolic rates or on alternative strategies for dissipation~\cite{speakman_maximal_2010}. The example of the elephant ears is emblematic, as these large ears increase the surface area--volume ratio of the animal and bring a net gain in heat dissipation~\cite{williams_heat_1990}. 

However, all mammals do not have ears as large as those of the elephants, but they all have an internal 
surface in contact with the ambient air: the lung. 
The decrease of the relative capacity to dissipate heat of the skin with the mammals' mass raises 
the question on the role of this other surface on the global dissipation strategy of mammals. The structure of the lung offers by far the largest exchange area of mammals with the outside world. Typically, in human, the mean surface area of the lung is about $75 \ \rm{m}^2$~\cite{weibel_pathway_1984}, while the surface of the skin is about $2 \ \rm{m}^2$~\cite{mosteller_simplified_1987}. The role of the lung, and actually of all the respiratory tract, is often neglected regarding the heat exchanges since its contribution is considered negligible~\cite{speakman_maximal_2010}. However, the respiratory tract has been estimated to dissipate up to $10 \%$ of the total heat produced in human~\cite{kirch_temperature_2005} with no decisive insight on how these $10 \%$ distribute between the upper respiratory tract~\footnote{The upper respiratory tract corresponds to the nasal cavity, the pharynx and the larynx.} and the lung. Heat Dissipation Limit theory~\cite{speakman_maximal_2010} indicates however that an increase of up to $10 \%$ in heat dissipation might be a decisive evolutionary advantage, especially for large animals. Also, mammals' lung characteristics are known to be finely adjusted to the animal's mass, with striking allometric scaling laws~\cite{west_general_1997, weibel_exercise-induced_2005}. Hence, the contribution of mammals' lung on heat dissipation is ought to depend on the animal's mass, potentially in the form of allometric scaling laws. The derivation and comparison of such laws with those of the metabolic rates could highlight the place of the lung in the general heat dissipation strategy of mammals. As complete data for allometric laws is available only at basal metabolic rate, we will focus our analyses on this regime using the framework proposed by West et al.~\cite{west_general_1997}. 

During inspiration, ambiant air is brought in contact with the mucosa of the mammals’ respiratory tract, which is usually warmer and wetter than the inhaled air. Hence, during inspiration, heat should be extracted from this tissue by direct exchanges with the air, i.e. conductive and convective transports, and by evaporation of the water contained in the mucosa~\footnote{phase change due to a water activity imbalance.}. It was first thought in the 1950s that the air was fully heated and humidified in the upper respiratory tract of human adults, i.e. before even reaching the lung~\cite{cole_recordings_1954, walker_heat_1961}. This was refuted in the 1980s by experiments conducted by McFadden et al.~\cite{mcfadden_respiratory_1983}. They noticed that air temperatures in the proximal bronchi were significantly below body temperature. Such an observation suggests that heat is extracted from the mucosa of the lung during inspiration, decreasing the mucosa temperature and inducing a cooling of the expiratory air~\cite{mcfadden_thermal_1985}. However, the intricate interplay between these transfers and the complex geometry of the lung is still not understood as of today: how and how much heat is really dissipated by the lung remain to be uncovered. More generally, this raises the question about the role and about the quantification of the lung as a heat and water exchanger within the body, for humans and, more generally, for mammals. 

The analysis proposed in this paper is based on an original, fully tractable and analytical mathematical model of the physical processes of heat and water exchanges in the lung. It is the result of a comprehensive simplification of an advanced, but less tractable, computational model that forms the most detailed validated modelling work to date~\cite{karamaoun_new_2018} and that reproduces qualitatively well the reference experiments from McFadden et al.~\cite{mcfadden_thermal_1985}. Our analytical model is validated by the strikingly fact that its predictions also reproduces qualitatively well the same experiments, with a degree of quality that is close to that of the advanced computational model.\\

\begin{figure*}[ht!]
(a)
\includegraphics[height=4.25cm]{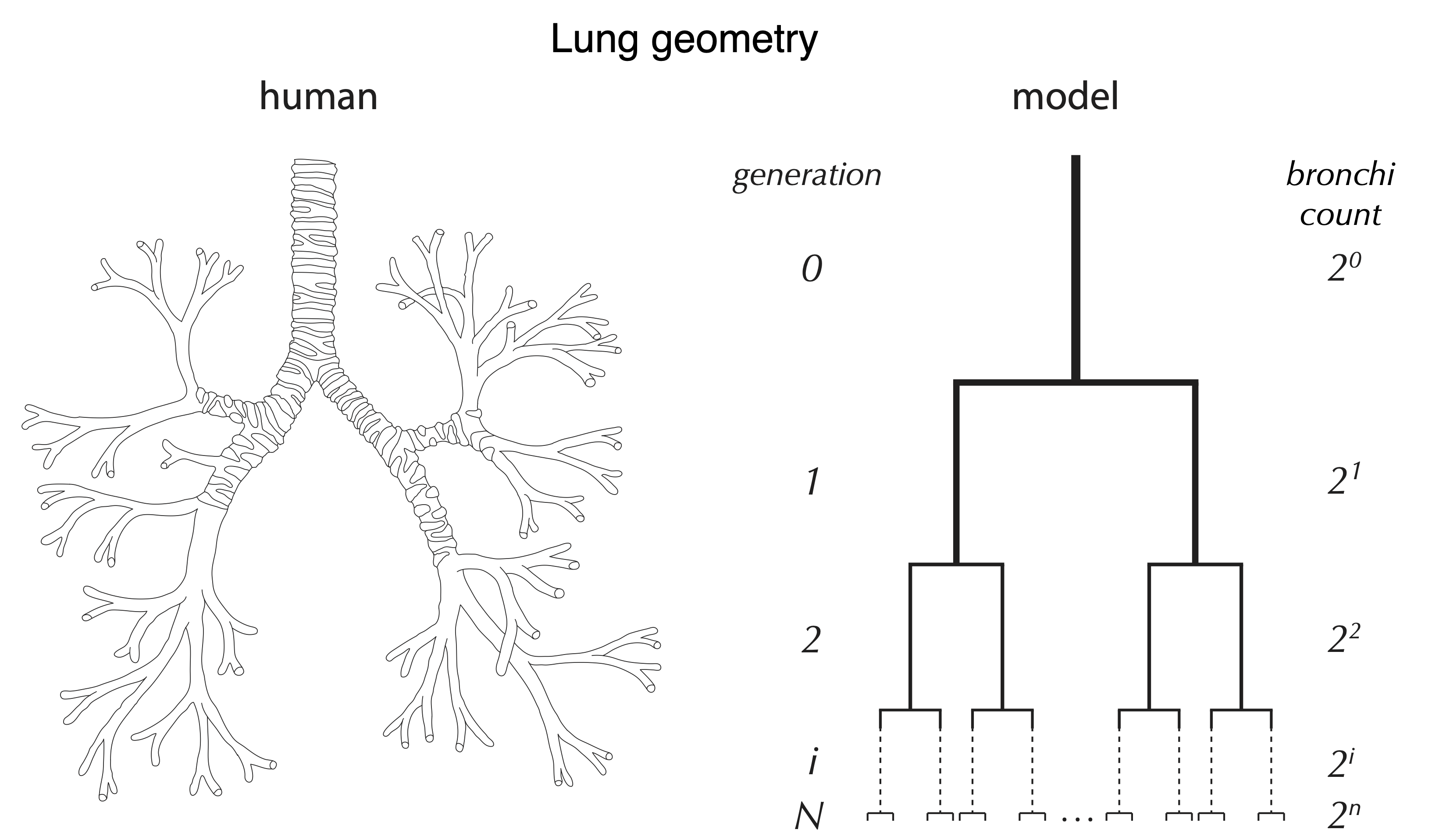}
(b)
\includegraphics[height=4.25cm]{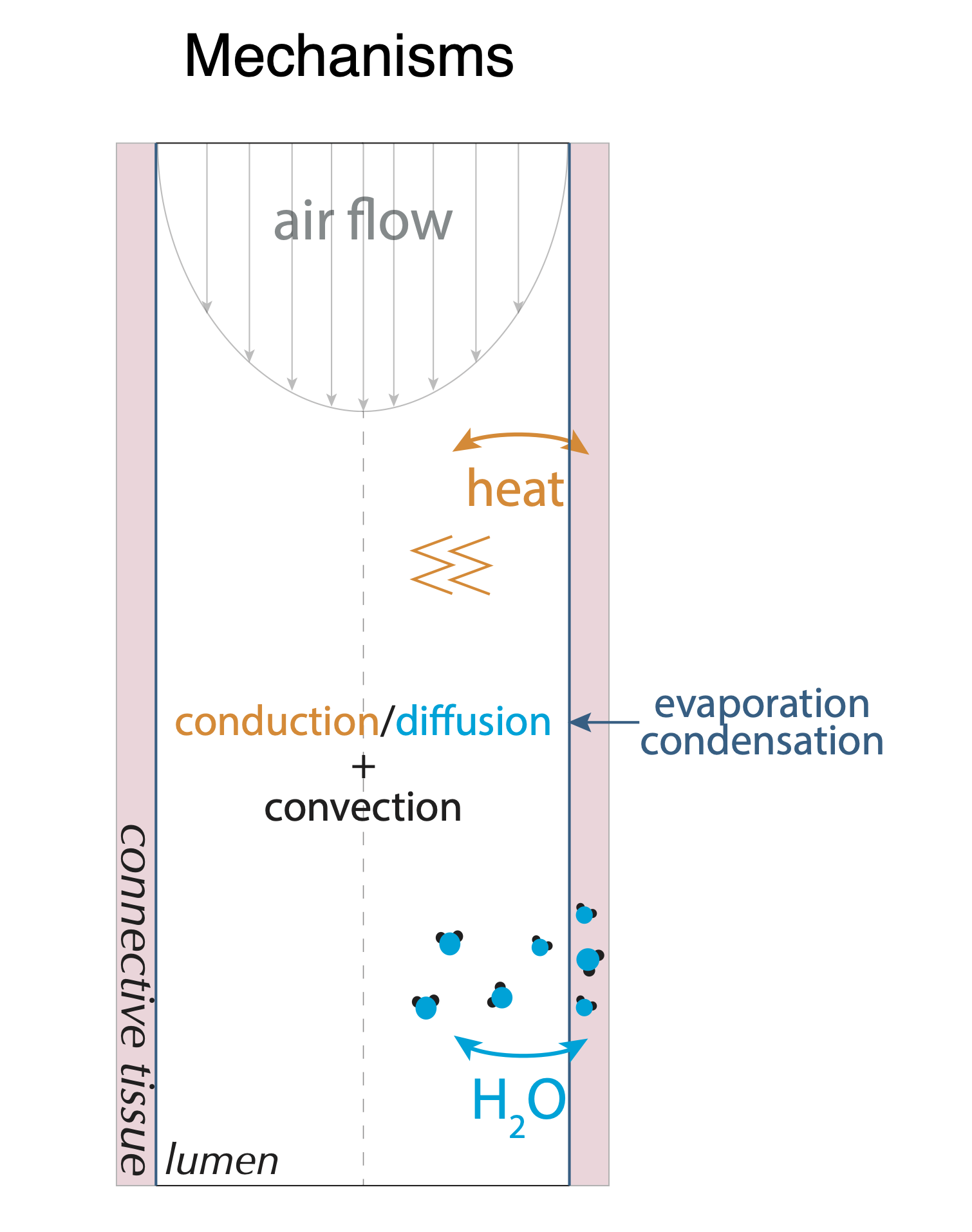}
(c)
\includegraphics[height=4.25cm]{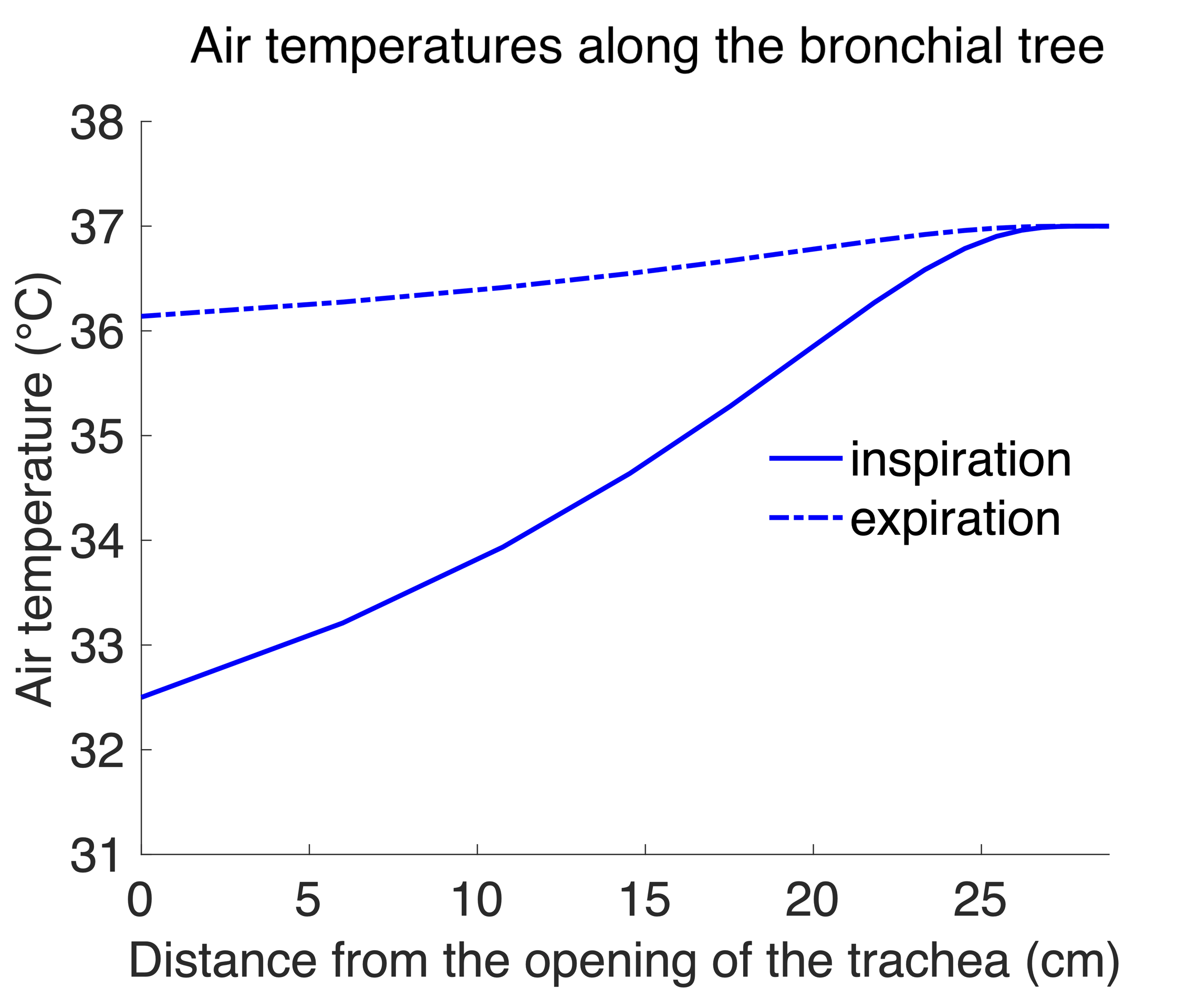}
\caption{(a) The mammals' bronchial tree (left) is modelled using Weibel "A" model~\cite{weibel_morphometry_1963} (right). (b) The power exchange is studied in axisymmetric idealized bronchi; it is mainly driven by the dynamics of water in the bronchi. Water is transported through the tree and in the bronchi by advection with air and diffusion. While transported, water condensates on or evaporates from the bronchi mucosa.  
(c) An example of the resulting air temperature predicted by our analytical model in the case of the human at rest. These predictions are in good agreement with the experimental data from McFadden et al.~\cite{mcfadden_respiratory_1983} and are very close to those from the computational model from Karamaoun et al.~\cite{karamaoun_new_2018}.}
\label{Figure1}
\end{figure*}

The Weibel "A" model has already proven to be very reliable to mimic the geometry of the mammals' lung for analyzing a wide range of lung-related processes~\cite{weibel_morphometry_1963, mauroy_optimal_2004, tawhai_lung_2009, mauroy_toward_2011, hofmann_modelling_2011, karamaoun_new_2018, noel_interplay_2019}. 
It consists in a symmetric bifurcating tree with straight circular cylinders mimicking the bronchi. The dimensions of the cylinders decrease with a factor $h = \h \simeq 0.79$ at bifurcations~\cite{weibel_morphometry_1963, mauroy_optimal_2004}. The set of bronchi with the same dimensions is called a generation; generations are numbered from $0$ to $N$. The first generation ($i=0$) corresponds to the trachea and the last generation ($i=N$) corresponds to all the bronchi that are connected to the alveolar regions (acini)~\footnote{The exact determination of the last generation index of the branching tree $N$ is not important in the framework of the presently developed analytical model as soon as $N$ is large enough, because saturation of air with water is obtained far before reaching the acini. However, generally, $N$ is chosen as the highest generation index $i$ satisfying $r_i > d_{\rm alv}$, where the alveolar duct diameter $d_{\rm alv}\simeq 0.141\times 10^{-3}\ M^{\frac{1}{12}}$.}. The $i^{\rm th}$ generation consists of $2^{i}$ bronchi with the same radius $r_i = h^{i} r_0$, with length $ l_i = h ^ {i} l_0 $, cross-sectional area $S_i = \pi r_i^2$  and area of the mucosa surface in contact with the air $W_i = 2 \pi r_i l_i$.  A schematic view of the model geometry is presented in Fig.~\ref{Figure1}(a). As shown in Table \ref{theTable}, core quantitative characteristics of the mammals' lungs are described by empirical or computed allometric scaling laws~\cite{west_general_1997, parent_comparative_2015}. 

The power exchanged between the bronchial mucosa and the air within the bronchial lumen is assumed to be strictly spent on water vaporization or condensation, and on air heating or cooling, depending on the conditions, as schematized on Fig.~\ref{Figure1}(b). Interestingly, the amount of power exchanged for heating or cooling the air represents a quasi constant fraction $\beta \simeq 15\%$ of the amount of power exchanged to vaporize or condense water~\footnote{The power exchanged between the bronchial mucosa and the breathed air is spent to vaporize/condense the water, and to heat/cool the air depending on the conditions. Thus, for instance at inspiration, the powers exchanged in a bronchus of generation $i$ writes $\Pin_i = v_i S_i \L  (\Cin_{i} - \Cin_{i-1})+ v_i S_i \rho_g c_{p,g}(T^{\rm{in}}_{i}-T^{\rm{in}}_{i-1})=(1+\beta)v_i S_i \L (C_i-C_{i-1})$ where $\Cin_{i}$ and $T^{\rm{in}}_{i}$ are respectively the water concentration and the air temperature in the generation $i$. The dimensionless number $\beta$ compares the quantity of energy required to heat/cool the air to the one required to phase change the water, it expresses as $\beta=\frac{\rho_g c_{p,g}}{\L}\frac{(T_b-T_0)}{ (C_{\rm sat}(T_b)-C_0)}$ and varies little within the range of temperatures explored by the system we are studying. $\beta$ can be estimated to about $0.15$. This number reveals that most of the power dissipated in a bronchus is used for phase changes ($87 \%$). 
}. 
Hence, the power exchanged between the mucosa and the air within a bronchus is proportional to the flow of water exchanged between the mucosa and the air in this bronchus, allowing us to focus on the water exchanges only. 

In a bronchus of generation $i$, the air flows with a velocity $v_i>0$ at the inspiration and $v_i<0$ at the expiration, and the conservation of air volume directly implies $v_i S_i = 2 v_{i+1} S_{i+1}$. The inspired air carries a water concentration $\Cin_i$ and the expired air carries a water concentration $\Cex_i$. The powers exchanged in a bronchus of generation $i$ at inspiration and expiration respectively express $\Pin_i = (1+\beta) \L v_i S_i (\Cin_{i} - \Cin_{i-1})$ and $\Pex_i = (1+\beta) \L v_i S_i (\Cex_{i-1} - \Cex_{i})$, with $\L$ the molar latent heat of water vaporization. The total power exchanged 
is then
$\P = \frac{1+\beta}{2} \L v_0 S_0 \left( (\Cin_N - \Cin_0) + (\Cex_0 - \Cex_N) \right)$. During the whole respiratory cycle, we assume that the air in the deepest region of the lung is fully heated and hydrated i.e. $\Cin_N = \Cex_N = 1$~\footnote{
The fact that distal air in the lung is fully heated and hydrated can be proven using an infinite tree and sending this condition to infinite, $\lim_{i \rightarrow \infty} \tCin_i = 1$ and $\lim_{i \rightarrow \infty} \tCex_i = 1$. Then our model predicts that for $N=17$, the shift of $\tCin_N$ and $\tCex_N$ from $1$ is less than $0.1 \%$. Hence this supports that assuming $\tCin_N = 1$ and $\tCex_N = 1$ is a reasonable approximation.}. The total power exchanged in the lung can be rewritten $\P = \frac{1+\beta}{2} \L v_0 S_0 (\Cex_0 - \Cin_0)$.
$\Cin_0$ depends on the heating and on the hydration of the air in the upper respiratory tract at inspiration, and thus on mammal species and on their environment. Nevertheless, in order to derive a coherent response between species, the same $\Cin_0$ 
will be applied for all mammals~\cite{Note4}. 
With this assumption, dimensionless water concentrations can be used: $\tCin_i = (\Cin_i - \Cin_0)/(C_{\rm{sat}}(T_b) - \Cin_0)$ and $\tCex_i = (\Cex_i - \Cin_0)/(C_{\rm{sat}}(T_b) - \Cin_0)$, where $C_{\rm{sat}}(T_b)$ is the saturation concentration of water in air at the mammal's body temperature $T_b$~\cite{clarke_scaling_2007}.
The corresponding normalized total power exchanged in the lung is then
\begin{equation}
\tP = \frac{1+\beta}{2}\L v_0 S_0 \tCex_0 \propto M^{\frac34} \tCex_0 
\label{power}
\end{equation}

During ventilation, the stationary balance of the amount of water in a bronchus at generation $i$ is equal to
the difference between the water convected by the airflow entering and leaving the bronchus, 
to which is added the water exchanged by phase change with the mucosa, as illustrated in Fig.~\ref{Figure1}(b). The exchange with the mucosa is driven by the difference between the water concentrations at the air--mucosa interface ($\tCmu_i$) and the water concentration in the lumen, approximated by its mean value $\tC_i$. At equilibrium, $\tCmu_i$ equals the saturation concentration of water $\tilde{C}_{\rm sat}(T_i)$ at the mucosa temperature $T_i$. Mathematically,
\begin{equation}
\underbrace{v_{i} S_{i} \tC_{i-1}}_{\text{inflow (if $v_{i}>0$)}} + \underbrace{u_i W_i ( \tCmu_i - \tC_i )}_{\text{exchange with mucosa}} = \underbrace{v_{i} S_{i} \tC_{i}}_{\text{outflow (if $v_{i}>0$)}}.
\label{waterBalance}
\end{equation}
The quantity $u_i = \sqrt{\frac{D |v_i|}{l_i}}$ is here the diffusion velocity of water in the bronchus during a transit time of air $t_{a,i}=l_i/|v_i| = 2h^3 l_0/ |v_0|$. Since $h = \h$, $t_{a,i}$ and $u_i$ do not depend on the generation and the index $i$ can be dropped. Equation (\ref{waterBalance}) can be reformulated during inspiration into $\tCin_i = \tCin_{i-1} + \Gamma_i (\tCmu_i - \tCin_i)$ and during expiration into $\tCex_i = \tCex_{i-1} - \Gamma_{i} (\tCmu_i - \tCex_{i})$. 
The quantity $\Gamma_i = 2 \frac{u}{r_i} \frac{l_i}{|v_i|}$ measures the strength of water transport in the air from the air--mucosa interface to the lumen of the generation $i$~\footnote{$\Gamma_i $ is twice the ratio between two characteristic times: $t_{a,i} = l_i / |v_i| = 2 h^3 l_0 / |v_0|$~s that represents the transit time of air due to ventilation in a bronchus of generation $i$ and $t_{d,i} = \frac{r_i}{u}= \frac{h^{i} r_0}{u}$~s that represents the time for water to diffuse towards the center of the lumen of the same bronchus. Note that $t_{a,i}$ is actually independent of the generation $i$ since $h$ is constant. Consequently, $\Gamma_i$ increases with the generations since the diffusion time decreases along the generations. Finally note that both characteristic times $t_a$ and $t_{d,i}$ are most of the time much smaller than the lung's ventilation characteristic time $t_{\rm{in}} \simeq 2$~s, supporting the quasi-static approximation of the water flow balance in equation~(\ref{waterBalance}).}. An allometric scaling analysis reveals (see Table~\ref{theTable}) that $\Gamma_i \simeq 0.83 \ 2^{\frac{i}{3}} \ M^{-\frac14}$~\cite{west_general_1997, parent_comparative_2015}. These transfer coefficients are smaller for large mammals because of geometrical effects only since their mass-dependence is driven by the bronchial aspect ratio $r_i/l_i \propto M^{\frac14}$ that increases with the animal's mass. 

\begin{table*}[t!]
\centering
\medskip
\begin{tabular}{ccccc}
\hline
Physiology & & Allometry (rest) & Units & References\\
\hline
$h$ & size reduction of bronchi at bifurcation & $(1/2)^{\frac13} \simeq 0.79\ {\color{black} M^{0}}$ & -- & \cite{weibel_morphometry_1963, mauroy_optimal_2004}\\
$r_0$ & tracheal radius & ${2.03\times 10^{-3}}\ M^{\frac38}$ & m & \cite{west_general_1997}\\
$l_0$ & tracheal length & ${2.07\times 10^{-2}}\ M^{\frac14}$ & m & \cite{tenney_comparative_1967, west_general_1997}\\
$v_0$ & tracheal mean air velocity (inspiration) & ${0.80}\ {\color{black}M^{0}}$ & m.s$^{-1}$ & \cite{west_general_1997}\\
$t_{a}$ & characteristic time of air transit in a bronchus & ${2.61\times 10^{-2}}\ M^{\frac14}$ & s & \cite{west_general_1997}\\
$\tin$ & characteristic time of inspiration/expiration & ${0.69}\ M^{\frac14}$ & s & \cite{west_general_1997}\\
$t_{b}$ & equivalent transit time of blood in the tissue & $8.49 \ M^{\frac14}$ \footnote{This quantity is a characteristic time of blood transport in a tissue and is the ratio between a blood flow and a volume of tissue. In general, flows are proportional to $M^{\frac34}$ and volumes to $M^1$~\cite{west_general_1997, lindstedt_pulmonary_1984}, so we assume that this transit time follows an allometric scaling law in $M^{\frac14}$. The allometric law is then determined from that of human~\cite{bui_modeling_1998}.} & s & \cite{west_general_1997, bui_modeling_1998, karamaoun_new_2018}\\
\\
\hline
Thermodynamics & & &Units\\
\hline
$\L$ & molar latent heat of water vaporisation & ${43.49}$ & kJ.mol$^{-1}$ & --\\
$D$ & diffusion coefficient of water vapor into air & ${2.705\times 10^{-5}}$ & m$^2$.s$^{-1}$ & --\\
$\alpha_l$ & thermal diffusivity of liquid water ($\sim$ tissue) & ${1.51\times10^{-7}}$ & m$^2$.s$^{-1}$ & --\\
$\rho_{l}$ & density of liquid water  & 993  & kg.m$^{-3}$ & --\\
$\rho_{b}$ & density of blood  & 1060  & kg.m$^{-3}$ & --\\
$c_{p,l}\approx c_{p,b}$ & heat capacities of liquid water and blood &  4178 & J.kg$^{-1}$.K$^{-1}$ & --\\
$\delta_c$ & characteristic thickness of mucosa affected by heat exchanges \footnote{The quantity $\delta_c = \sqrt{\alpha_l \tin}$ is the characteristic diffusive exploration length of heat in the mucosa on the time scale of lung's ventilation $\tin$.} & $\sqrt{\alpha_l \tin}$ & m & \cite{karamaoun_new_2018}\\
\hline
\end{tabular}
\caption{Model's parameters and allometric scaling laws ~\cite{west_general_1997}. The thermodynamics quantities are given at 37$^{\circ}$C.}
\label{theTable}
\end{table*}

\begin{figure}
\noindent(a) \includegraphics[width=7cm]{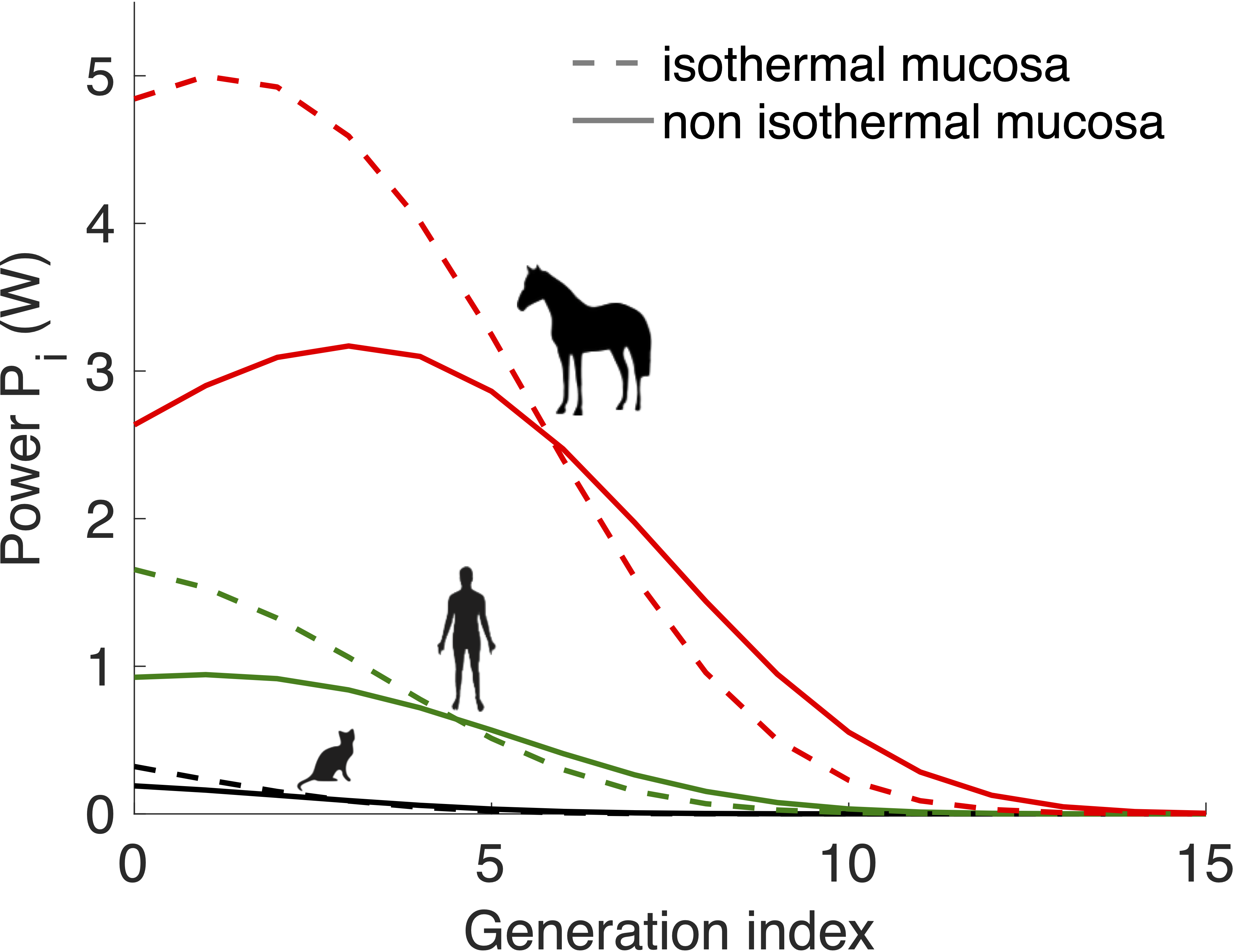}
\newline
\par \vspace{0.3cm}
\hspace{-0.7cm}(b) \includegraphics[width=7cm]{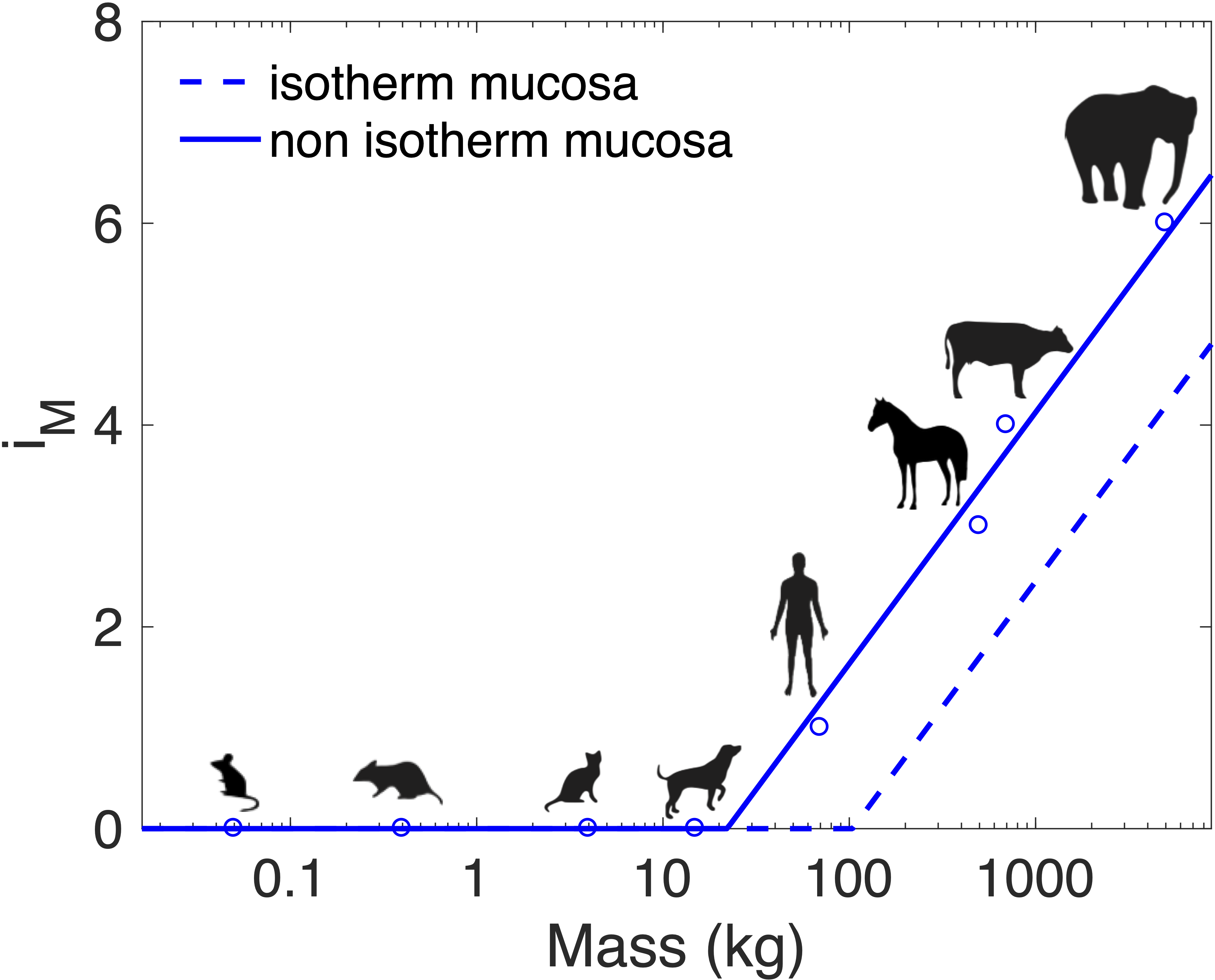}
\caption{(a) The distribution of the local dissipated power at rest along the generations of mammals' lungs is complex and the location of the maximal dissipation in the bronchial tree depends on the species. (b) The location of the maximum of the local dissipated power along the bronchial tree is proportional to $\frac34 \frac{\log(M)}{\log(2)}$.}
\label{Figure2}
\end{figure}

First, we will focus on a simpler inspiratory context by assuming that the air is at the body temperature at the air--mucosa interface, i.e. $\tCmu_i = 1$. 
Under this assumption, 
the power exchanged in the generation $i$ at inspiration, $\tp_i = 2^{i} \tPin_{i}$ follows a recurrence relation
$
\tp_{i} = \frac{2h^2}{\left( 1+\left( 2 h^2 \right)^{i} \Gamma_0 \right)} \tp_{i-1}
$.
Thus, the behavior of the local dissipated power $\tp_i$ along the tree generations results from a trade-off 
between the increase of the cumulated surface area of the bronchial sections along the generations (with the factor $2h^2$) and the decrease of the water concentration difference between the air--mucosa interface and the bronchus (with the factor $1/(1 + (2 h^2)^i \ \Gamma_0)$).
When $\Gamma_0$ is smaller than $1-h \simeq 0.21$, $\tp_i$ first increases along the bronchial tree up to a maximum value from where it starts to decrease. The maximum is reached at a generation $i_M = \left[1+\frac{\log\left(\Gamma_0 / (1-h) \right)}{\log(h)} \right]$~\footnote{In the isothermal case,
$$
i_{M,\rm{iso}} = \left[1+\frac{\log\left(\Gamma_0 / (1-h) \right)}{\log(h)} \right]
$$
while in the non isothermal case, a correction term is added due to the modification of exchanges because of the low mucosa temperature,
$$
i_{M,\rm{non iso}} = \left[ 1+\frac{\log\left( \Gamma_0 / (1-h) \right)}{\log(h)}-\frac{\log\left( \sqrt{1+\G} \right)}{\log(h)} \right]
$$
with $\G$ a universal number amongst mammals that we uncover in this paper, see the next steps of our analysis and \cite{Note16}.
}. When $\Gamma_0$ is larger than $1-h$, the maximum is reached at the trachea and $\tp_i$ decreases monotonically along the bronchial tree. The transition occurs in the physiological range of mammals' masses: the heat dissipation in small mammals is indeed maximal at the trachea, while for larger ones, it is maximal deeper in the bronchial tree, with the rule of the heavier, the deeper, as shown in Fig.~\ref{Figure2}~\footnote{As $\Gamma_0$ is increased at exercise, the location of the dissipation maximum shifts towards the proximal part of the mammals' bronchial tree, and the transition occurs for a larger mass than at rest. The location of the maximal dissipation might be related to the panting behavior, that is used by small mammals, dogs and sheeps~\cite{terrien_behavioral_2011}. Indeed, quick shallow breaths decrease $\Gamma_0$ and shift the maximum dissipation toward the proximal part of the tree, hence potentially increasing panting's efficiency~\cite{robertshaw_mechanisms_2006}.}. 

The simplified approach that assumes an isothermal mucosa leads to a behavior, highlighted in the previous analysis, which still holds with the non isothermal case. The analytical expression for $i_M$ in the frame of a non isothermal mucosa shows a shift toward the deeper bronchi relatively to the isothermal case, as shown by the next steps of our study and by \cite{Note16}. 
However, both models predict exactly the same slope for $i_M$ as a function of $\log(M)$: $\frac34 \frac{\log(M)}{\log(2)}$. 
The isothermal mucosa simplification highlights the complex competition between physics and geometry, but does not reproduce the drop in temperature of the expired air observed by McFadden et al.~\cite{mcfadden_thermal_1985}. 

We will now relax the assumption of an isothermal mucosa and consider the mean temperature $T_i$ of the mucosa of a bronchus in the generation $i$. 
As for water concentration, a dimensionless temperature for the mucosa $\tT_i$ is defined relatively to the tracheal temperature of air at inspiration $T_0^{\rm{in}}$: $\tT_i = (T_i - T_0^{\rm{in}})/(T_{b} - T_0^{\rm{in}})$. 
At inspiration, energy is lost from the bronchial mucosa by direct heat exchanges and water phase change in the lumen, while it is also reheated in parallel by the blood flowing in the connective tissue~\cite{weibel_pathway_1984}\cb . Thus, the deviation of the mucosa temperature from the body temperature depends on the relative strength of these two phenomena, and formulates using an energy balance equation~\cite{karamaoun_new_2018}~\footnote{
According to the assumptions stated and discussed in~\cite{karamaoun_new_2018}, the energy balance equation for the mucosa in a bronchus in generation $i$ can be written
\begin{multline}
c_{p,l} W_i \sqrt{\alpha_l \tin}\frac{dT_{i}}{dt} =\\
W_i \sqrt{\alpha_l \tin} \frac{c_{p,b} Q_b}{V_c} (T_b - T_{i})\\
 -(1+\beta) \L W_i k (\tCmu_i-\tilde{C}_i)
\end{multline}
The quantity $W_i \sqrt{\alpha_l \tin}$ represents the characteristic volume of the mucosa ring around the bronchus affected by the exchanges with the air in the lumen. The thickness of the ring is assumed small relatively to the radius of the bronchus, typically about $0.5$ mm in human. Hence, the region where the tissue temperature is affected stands within the bronchi walls~\cite{montaudon_assessment_2007}. The first term of the right handside corresponds to the heating of the mucosa from the blood flow, while the second term corresponds to the cooling or heating of the mucosa by heat exchanges at the air--mucosa interface and by water phase change at the same interface.
$Q_b$ is the blood flow rate in the bronchial circulation, $V_c$ the volume of the connective tissue around the bronchi in the lung. Equivalent transit time of blood is then $t_b = V_c/Q_b$. The quantity $k$ is a mass transfer coefficient between the air--mucosa interface and the core of the lumen. According to the classical boundary layer theory, $k$ is expressed as $k = D/\sqrt{D t_a}$, with $t_a$ the transit time of air in the bronchus, independent of $i$. The quantity $\delta_{c} =  \sqrt{\alpha_l \tin}$ is the characteristic depth explored by the heat diffusion in the mucosa during the inspiration time $\tin$. This equation is then normalised in time with $\ttt = t/\tin$.
},
\begin{equation}
\frac{d \tT_{i}}{d\ttt} =
\underbrace{\Xi (1 - \tT_{i})}_{\footnotesize \begin{array}{c}\text{heating from }\\ \text{blood flow}\\ \text{in the mucosa}\end{array}} - \underbrace{\Psi \left(\tCmu_i - \tC_i\right)}_{\footnotesize \begin{array}{c}\text{cooling/heating from}\\ \text{direct heat exchanges}\\ \text{and phase change}\\ \text{in the lumen}\end{array}}
\label{Tconj}
\end{equation}

\noindent  where
$
\Xi = \beta_{\Xi} \ \frac{\tin}{t_{b}} \ \text{ and } \ \Psi = \beta_{\psi} \  \sqrt{\frac{\tin}{t_a}}
$
are two dimensionless numbers.
The number $\Xi$ characterizes the strength of the mucosa reheating by the renewal of the blood flowing in the connective tissue. 
It is proportional to the ratio of two physiological characteristic times: the characteristic time of inspiration $\tin$ and 
the transit time of blood in a volume equivalent to the whole volume of the bronchial mucosa $t_{b}$.
The dimensionless coefficient $\beta_{\Xi} \simeq 1.07$~\footnote{$\beta_{\Xi} = \frac{\rho_b c_{p,b}}{\rho_l c_{p,l}}$.}
is the ratio of the volumetric heat capacities of the blood and water that balances $\tin$ and $t_b$ accordingly.
The number $\Xi$ is intrinsically independent of the generation within the bronchial tree. The same is true for the number $\Psi$ since $t_a$ was shown to be independent of $i$. This number characterizes the strength of the cooling of the mucosa by heat and water exchanges from the mucosa interface to the lumen where air flows.
It also depends on the ratio of two physiological characteristic times: the characteristic time of inspiration $\tin$ and the characteristic time of air transit in a bronchus $t_{c}$. The dimensionless coefficient $\beta_{\Psi} \simeq 0.025$ 
balances $\tin$ and $t_c$ accordingly to the diffusive properties of heat and the diffusive and evaporative properties of water in air and mucosa~\footnote{$\beta_{\Psi} = (1+\beta) \rm{Ko} \ {\rm{Le}}^{-\frac12} \simeq 0.025$ is composed of two dimensionless numbers: the Kossovitch number ${\rm{Ko}}=\frac{\L (C_{\rm sat}(T_b) - \Cin_0)}{ \rho_l c_{p,l} (T_b - T_0^{\rm in})} \simeq 1.62\times10^{-3}$ 
which compares for a wet medium the energy required for the phase change to the one required for its heating or cooling, and the Lewis number ${\rm{Le}}=\alpha_l / D \simeq 5.58\times 10^{-3}$ that compares the heat diffusion in the mucosa with the vapor diffusion in the lumen. Its estimated value is computed for $T_b = 38^{\circ} C$.}.   
The allometric scaling laws of the physiological characteristics involved in the expressions $\Xi$ and $\Psi$ (see Table~\ref{theTable}) show that neither these two numbers depend on mammals' mass with $\Xi \simeq 0.087$ and $\Psi \simeq 0.15$. 
Hence, the thermophysical response of the bronchial mucosa to temperature and water concentration variations is identical all along the bronchial tree of a single species, as well as for all the Mammalia class.

Estimates of $\Xi$ and $\Psi$ indicate that both the dynamics of heat supply by the blood flow within the mucosa and heat transfer with the air flow in the lumen are of similar order of magnitude, but still relatively small in front of $1$. Thus, the mucosa temperature varies relatively slowly over a ventilatory cycle. The water concentration $\tCmu_i$ at the air--mucosa interface is driven by the mucosa temperature and can be considered independent on time and approximated with $\tCmu_i \simeq \frac{2 + \frac{\Upsilon \Psi}{\Xi} (\tCin_i + \tCex_i)}{2 + 2 \frac{\Upsilon \Psi}{\Xi}}$\footnote{If we assume a local equilibrium at the air--mucosa interface and small shifts of mucosa temperature relatively to body temperature, an approximation of water concentration at air--mucosa interface can be derived, $\Cmu_i(t) = C_{\rm sat}(T_i) \simeq C_{\rm sat}(T_b) + \frac{d C_{\rm sat}}{dT}(T_b) \times (T_i(t) - T_b)$ or, in dimensionless form, $\tCmu_i(\ttt) =  1 - \Upsilon (1 - \tT_i(\ttt))$, with $\Upsilon = \frac{d C_{\rm sat}}{dT}(T_b) \frac{T_b - T_T}{C_{\rm sat}(T_b) - \Cin_0}$.
Formulated for inspiration (exponent $^{\rm{in}}$) and expiration (exponent $^{\rm{ex}}$), the equation (\ref{Tconj}) can be integrated on dimensionless time in $[0,1]$, 
\begin{equation}
\begin{array}{l}
\tTex_i(1) - \tTex_i(0) =\\
\hspace{1cm} \Xi/\Upsilon \ (1 - \tCmu_{i,\rm{ex}}) - \Psi (\tCmu_{i,\rm{ex}} - \tCex_i)\\
\\
\tTin_i(1) - \tTin_i(0) =\\
\hspace{1cm} \Xi/\Upsilon \ (1 - \tCmu_{i,\rm{in}}) - \Psi (\tCmu_{i,\rm{in}} - \tCin_i)
\end{array}
\label{linearEq}
\end{equation}
with the quantities on the right hand side being the mean values over the dimensionless time in $[0,1]$. Accounting for the slow time-variation of the mucosa temperature due to the fact that $\Xi$ and $\Psi$ are small, the mean temperatures can be approximated with
$\tTex_i \simeq \frac12 (\tTex(1) + \tTex(0))$ and $\tTin_i \simeq \frac12 (\tTin(1) + \tTin(0))$ (trapezoidal approximation). As end inspiration phase is the beginning of expiration phase and vice-versa, $\tTin_i(1) = \tTex_i(0)$ and $\tTex_i(1) = \tTin_i(0)$. Hence, $\tCmu_{i,\rm{ex}} - \tCmu_{i,\rm{in}} = \Upsilon (\tTex_i - \tTin_i) \simeq 0$ (at the order of the trapezoidal integration), and $\tCmu_{i,\rm{ex}} \simeq \tCmu_{i,\rm{in}}
 \simeq \tCmu_{i}$. This last equality allows to extract from the integrated equations (\ref{linearEq}) the approximation proposed for $\tCmu_i = (2 + \frac{\Upsilon \Psi}{\Xi} (\tCin_i + \tCex_i))/(2 + 2 \frac{\Upsilon \Psi}{\Xi})$.
 
This analysis validates the use of mean values for water concentrations for inspiration and expiration, most particularly in equation (\ref{waterBalance}).}. A third dimensionless number emerges $\Upsilon \simeq 0.80$~\footnote{$\Upsilon = \left. \frac{d C_{\rm sat}}{dT} \right|_{T_b} \frac{T_b - T_0^{\rm in}}{C_{\rm sat}(T_b) - \Cin_0}$, its estimated value is computed for $T_b = 38^{\circ} C$.}, which depends on constants related to water thermodynamics at $T_b$ and inhaled air properties, but $\Upsilon$ does not depend on the animal's mass either.
Interestingly, the three dimensionless numbers that appear in the expression of $\tCmu_i$ are grouped into a global dimensionless number $\G = \frac{\Upsilon \Psi}{\Xi} = \beta_{\G} \frac{t_b}{\sqrt{\tin t_a}} \simeq 1.35$. $\G$ is fully independent of the inhaled air properties, of the body temperature and of the mammals' mass. The dimensionless coefficient $\beta_{\G} \simeq 0.019$ is only composed of thermodynamical constants~\footnote{$\G = \frac{\Upsilon \Psi}{\Xi} = \beta_{\G}\frac{t_b}{\sqrt{t_a \tin}}$ with $\beta_{\G}=\frac{\L}{\rho_b c_{p,b}} \left.\frac{d C_{\rm sat}}{d T} \right|_{T_b} \rm{Le}^{-\frac{1}{2}}$. The quantity $\left.\frac{d C_{\rm sat}}{d T}\right|_{T_b}$ varies very little with $T_b$, hence it can be considered almost constant on the range of body temperature of mammals.}. 


The concentrations of water throughout the lung with a non isothermal mucosa have a striking analytical expression~\footnote{First, we eliminate $\Cmu_i$ and reformulate the equations in matrix form: $(Id + \Gamma_i A) c_i + \Gamma_i b = c_{i-1}$, with
$
c_i = \left(\begin{array}{c} \tCin_i \\ \tCex_i \end{array} \right)
$, $b = \frac{1}{1+\G}\left(\begin{array}{c} -1 \\ 1 \end{array} \right)$, $A= \frac{1}{2(1+\G)}
\left(
\begin{array}{cc}
2+\G & -\G\\
\G & -(2+\G)
\end{array}
\right)$ and $Id= \left(
\begin{array}{cc}
1 & 0\\
0 & 1
\end{array}
\right)$. The analytical solution is computed thanks to two observations: 1/ the quantity $\frac{c_{i-1} - c_i}{\Gamma_i} - A c_i = b$ is independent on $i$; 2/ the quantity $d_i = \frac{c_{i-1} - c_i}{\Gamma_i}$ is actually an eigenvector of $A$. Several algebraic manipulations and the assumption that $\tCin_0 = 0$ and $\tCex_N = 1$ (possibly with $N = +\infty$), allows to reach an analytical expression for the dimensionless concentrations,
$$
\begin{array}{l}
\tCin_i = \frac{\sum_{j=1}^i \Gamma_j  \prod_{k=1}^j \frac{1}{1 + \lambda \Gamma_k}}{\sum_{j=1}^N \Gamma_j \prod_{k=1}^{j} \frac{1}{1+\lambda \Gamma_k}}\\
\tCex_i = 1 - \frac{\sum_{j=i+1}^N \Gamma_j  \prod_{k=1}^j \frac{1}{1 + \lambda \Gamma_k}}{\sum_{j=1}^N \Gamma_j \prod_{k=1}^{j} \frac{1}{1+\lambda \Gamma_k}} \left( 1 + \frac{2}{\G} \left( 1 - \sqrt{1 + \G}\right) \right)\\
\lambda = 1/\sqrt{1+\G}
\end{array}
$$
From these equations, the location of the maximum of the dissipated power can be derived by solving the inequality $\tp_{i-1} < \tp_i$ relatively to $i$, with $\tp_i = \L 2^i v_i S_i (\tCin_i - \tCin_{i-1} + \tCex_{i-1} - \tCex_i)$, the power dissipated in generation $i$. Then,
$$
i_{M,\rm{non iso}} = \left[ 1+\frac{\log\left( \Gamma_0 / (1-h) \right)}{\log(h)}-\frac{\log\left( \sqrt{1+\G} \right)}{\log(h)} \right]
$$}.
The expiratory water concentration in the trachea is a function of $\G$ only and does not depend explicitly on the local exchanges described by $\Gamma_i$,
$$
\tCex_0 = \frac{2}{\G} \left(\sqrt{1+\G} - 1 \right) \simeq 0.81
$$
From water concentrations, temperatures can be derived, see~\cite{Note4}. Fig.~\ref{Figure1}(c) shows the predictions from our model of temperatures variation along the bronchial tree for human at rest. Even though the model of the geometry of the bronchial tree is not adjusted specifically to the subjects of the experiments of McFadden et al., the temperatures by our analytical model are in good qualitative agreement with their observations~\cite{mcfadden_thermal_1985}, see Fig.~\ref{Figure1}(c). Our model reproduces also very well the predictions of the advanced computational model from Karamaoun et al.~\cite{karamaoun_new_2018}.

From Eq.~$(\ref{power})$, the law describing the total normalized dissipated heat in the lung at rest as a function of the mammals' mass can finally be derived,
$$
\tP = \frac{1+\beta}{\G} \left(\sqrt{1+\G} - 1 \right) \L v_0 S_0 \simeq {0.21 \ M^{\frac34}} 
$$ 
The lung's heat exchange capacity is increased for small values of $\G$ and is maximal for $\G = 0$, when blood flow transfers a sufficient heat flow to maintain the mucosa at body temperature. Then, no energy is recovered during expiration, $\tCex_0 = 1$ and the maximal power $(1+\beta) \L v_0 S_0 / 2$ is dissipated. But in the physiological case $\G$ is strictly positive and the maximal power is weighted by the mass-independent ratio $\tCex_0 = \frac{2}{\G} \left(\sqrt{1+\G} - 1 \right) < 1$, that reflects the complex interactions between the physics and the lung's geometry. For large values of $\G$, the mucosa temperature drops closer to the inhaled air temperature and more energy is recovered during expiration. Consequently, the power dissipated in the lung decreases. In the limit case of $\G \rightarrow \infty$, no power is dissipated at all ($\tCex_0 \rightarrow 0$). Strikingly, mammals share the same universal value for $\G$ at rest, estimated to be about $1.35$. This suggests that under the same environmental conditions, their lungs should behave identically in term of heat and water exchanges, whatever the mammal's mass. 

The allometric scaling in $M^{\frac34}$ of the power $\tP$ is surprising considering that the dissipated heat patterns along the lung's generations can be fundamentally different between mammals, as shown on Fig.~\ref{Figure2}(a). The explanation holds on two underlying processes. First, the inspired air is quasi fully heated and hydrated when it reaches the deepest parts of the lung~\cite{Note5}, so the way the exchanges are occurring on its path does not influence the total power exchanged during inspiration, $\tPin = \frac{1+\beta}{2} \L v_0 S_0$.
Secondly, in each bronchus the expiratory air flow in contact with a colder mucosa releases a power that is proportional to the power dissipated during inspiration in the same place. But the ratio between the powers dissipated at expiration and inspiration in a bronchus is independent on both the generation index and the mammal's mass: {$|\tPex_i/\tPin_i| = 1 + \frac{2}{\G} \left( 1 - \sqrt{1 + \G}\right) \simeq 19 \%$}. The total power at expiration $\tPex$ is a fraction of the total power at inspiration $\tPin$, hence it is also quasi independent on the $\Gamma_i$~\footnote{The ratio $|\tPex_i/\tPin_i|$ is actually not constant in the advanced computational model from Karamaoun et al.~\cite{karamaoun_new_2018}, although the refined model also predicts a $3/4$ allometric scaling law for the total heat dissipated by the lung. Actually, the ratio $|\tPex_i/\tPin_i|$ shows a decrease that is slow in the upper bronchial tree and that is accelerating in the lower bronchial tree. Because most of the exchanges happen in the upper bronchial tree, the approximation of a constant $|\tPex_i/\tPin_i|$ still hold the core explanation of why the dissipated heat follows a $3/4$ allometric scaling law.}.

Our results confirm not only the hypothesis of McFadden about the role, in human, of the mucosa temperature on the temperature of the tracheal expired air~\cite{mcfadden_thermal_1985}, but also allow to quantify for the first time, with the universal number $\G$, the respective role of the physical phenomena involved in heat exchanges, for any mammal.\\

\begin{figure}
\includegraphics[width=7cm]{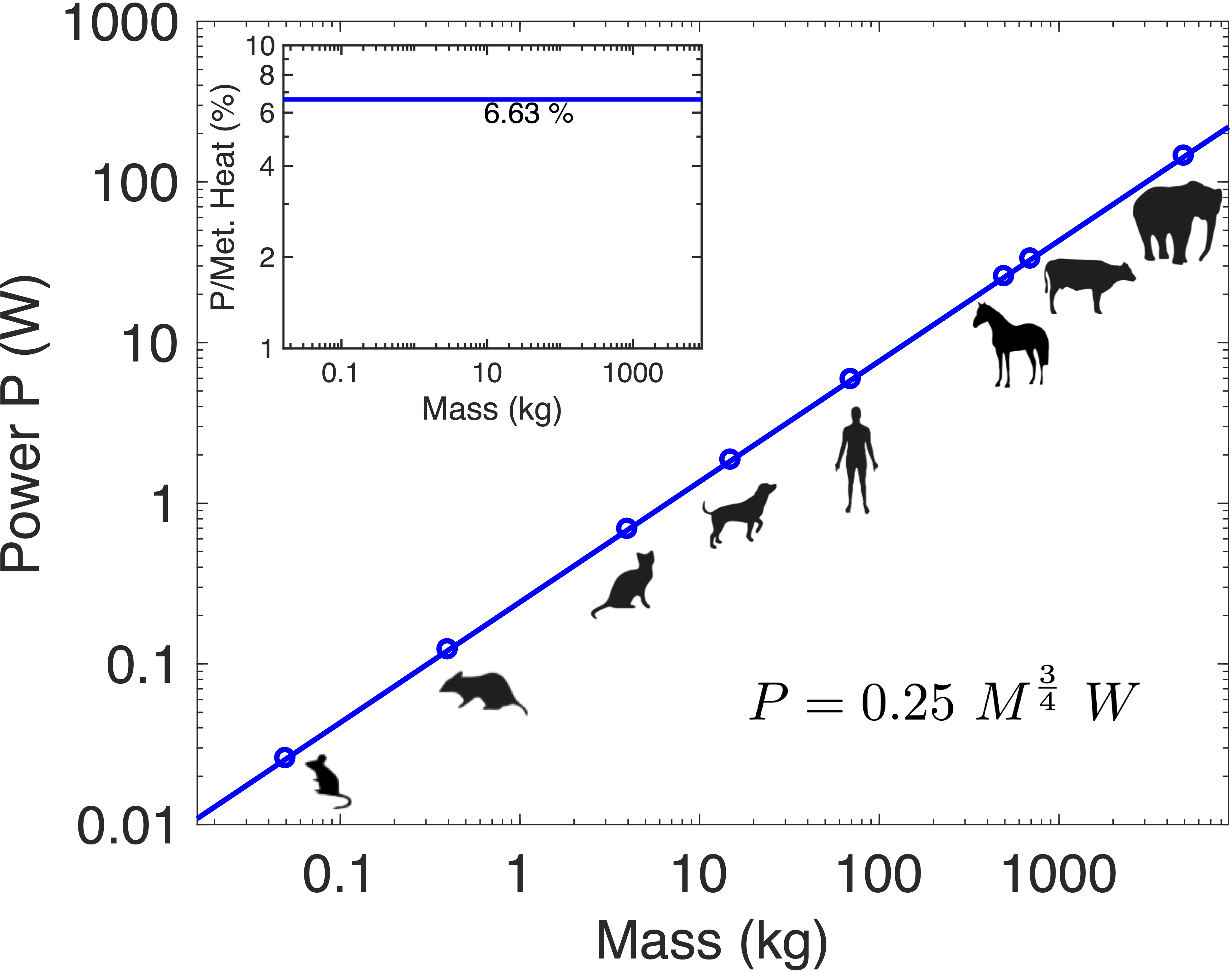}
\caption{Even though the distribution of the dissipated power in the lung is complex and depends on the mammal species, the total dissipated power by the lung at rest follows an allometric scaling law with an exponent ${\frac34}$, like the basal metabolic rate. Hence, whatever its size, a mammal dissipates with its lung the same proportion of the heat that it produces, about $6.6 \%$. Actually, the total dissipated power only depends on the universal dimensionless number $\G$, estimated to be  $\simeq 1.35$ at rest.}
\label{Figure3}
\end{figure}

{\bf Heat dissipation strategy.}
As heat production in the mammals' body is generally assumed to be proportional to $M^{\frac34}$~\cite{kleiber_body_1932}, the ratio between the heat dissipation and the heat production does not depend on the mammals' mass. Hence, all mammals dissipate the same proportion of heat under similar environmental conditions, as shown on Fig.~\ref{Figure3}. We estimated that 
$$
\P=0.25 \ M^{\frac34} \ \rm{W}
$$ 
over seven orders of magnitude in mammals' mass. It predicts that mammal's lungs are able to dissipate powers from 0.025 W for a mouse ($M=0.05$ kg), to 147 W for an elephant ($M=5000$ kg), through 5.98 W for a human ($M=70$ kg). Our model estimates that, in typical environmental conditions~\footnote{Considering water vapor as a perfect gas, the water concentration of air is linked to the temperature according to $C=P_{v}/(\hat{R}T)$, where $P_{v}$ is the vapor partial pressure and $\hat{R}=8.314$~J.mol$^{-1}$.K$^{-1}$ is the perfect gas constant. We classically consider that air reaches the trachea of the lung at inspiration with a humidity $H_0^{\rm in}=0.80$ and a temperature $T_0^{\rm in}=30^{\circ}$C \cite{mcfadden_respiratory_1983}. The water concentration of air at the trachea is then $C_0^{\rm in}=1.352$ mol.m$^{-3}$ since $P_v=H P_{\rm sat}(T)$, where $P_{\rm sat}(T)$ is the water saturation pressure estimated using the Clausius-Clapeyron equation. These values of $T_0^{\rm in}$ and $C_0^{\rm in}$ are considered in this work as input parameters for all mammals. The saturation concentration of water $C_{\rm sat}(T)$ estimates considering $P_v=P_{\rm sat}(T)$, and for instance equals to 2.436 at $T=T_b=37^{\circ}$C. Finally, $\left. \frac{d C_{\rm sat}}{dT}\right|_{T_b}=0.124145$.}, the lung dissipates about $6.6 \%$ of the total heat produced by the mammals' body at rest, while fulfilling its primary function, a number of the same order of magnitude than the $10 \%$ referenced in the literature~\cite{kirch_temperature_2005}. As dissipation through the skin is relatively less efficient in large mammals, the proportion of the heat dissipated by the lung versus that dissipated by the skin increases with the animals' mass with $M^{0.75}/M^{0.63} = M^{0.12}$. Hence, the relative role of the lung in the strategy of mammals for heat dissipation is more important for large mammals. 

The evolutive forces that scale the mammals' lung for respiration~\cite{mauroy_viscosity_2014} also scale the lung for heat dissipation. This result is not straightforward. Actually, even though respiratory and heat dissipation functions rely on transport and exchange phenomena, heat dissipation mainly occurs by water exchanges through the proximal bronchi walls, while oxygen and carbone dioxide exchanges occur in the distal regions of the lung where virtually no heat is exchanged. The evaporative cooling effect resulting in a mucosa temperature below body temperature is a key factor to understand the mammals' lung water exchanges and heat dissipation. 
At the individual level, mechanisms for blood flow regulation can affect the heat exchanges. Well known examples are the vasoconstriction of the pulmonary vessels or smooth muscles responses to cold environment~\cite{regnard_cold_1992}; both are suspected to be at the origin of some forms of asthma when dysfunctioning~\cite{shoraka_relation_2019}. 
At the species level, a trade-off between heat and water losses might well have been selected by evolution~\cite{rozenrechels_when_2019}. For example in human, our model predicts that the $6.6 \%$ of heat dissipated through the lung correspond to a daily loss of about $180$ ml of water, in full agreement with the usual estimations~\cite{Dmitrieva_increased_2011}.
From the analysis of the number $\G$, our model highlights the manner with which morphological parameters affect the heat dissipation.
The two characteristic times $\tin$ and $t_a$ are strongly related with the respiratory function of the lung. To the contrary, equivalent blood transit time in the connective tissue $t_b$ is a good candidate as an independent parameter that evolution might have acted on to respond to a potential trade-off between heat and water losses. Indeed, our work shows that heat dissipation through the lung has not been counter-selected. Furthermore, it might even have played a role in the evolution of the high metabolism of mammals~\cite{crompton_evolution_1978, speakman_maximal_2010}. This is all the more true for large mammals, as heat dissipation through their skin has a lower relative efficiency than small mammals~\cite{speakman_maximal_2010}.\\ 

{\bf Pulmonary diffusivities.}
A mammal with body temperature $T_b$ and with an inspiratory water concentration in the trachea $\Cin_0$ dissipates the absolute power $\P = \tP \times \left( \Csat(T_b) - \Cin_0 \right)$.
The quantity $(\Csat(T_b) - \Cin_0)$ represents the environmental conditions felt by the animal's lung, in the form of a water concentration difference in air between the trachea at inspiration ($\Cin_0$) and the distal lung, where the air is assumed to be saturated in water ($\Csat(T_b)$). 
The phenomenon of water exchanges can actually be reformulated at the whole organ level as an equivalent Fick's law,
$
\ja = -\D \left(\frac{\Csat(T_b) - \Cin_0}{l}\right)
$
where $\ja$ is the molar flux of water getting out of the lung and $l$ is the characteristic length between the trachea and the distal lung~\footnote{The length from trachea to the acini is $l = \sum_{i=0}^N l_0 h^i$ with $N+1$ the number of convective generations in the bronchial tree, then $l = l_0 (1-h^{N+1})/(1-h)$ and we assume that $N$ is large enough so that $1 - h^{N+1} \simeq 1$.}, $l = l_0/(1-h) \propto M^{1/4}$~\cite{stahl_scaling_1967, gunther_dimensional_1975}. The quantity $\D = \frac{v_0 l}{\G} \left(\sqrt{1+\G} - 1 \right) \propto M^{1/4}$ can be considered as an equivalent diffusion coefficient of water along the bronchial tree~\footnote{This formulation as a diffusive process allows to characterize the heat and water transport easily. Hence, the quantity $\vh = \D/l = \frac{v_0}{\G} \left(\sqrt{1+\G} - 1 \right) \propto M^0$ is the characteristic velocity of the water diffusive process throughout the lung, it is independent on the animal mass. The complex interactions between the lung geometry and the physics of exchanges imply that the actual diffusive velocity is smaller than the convective velocity, as it is a fraction $\frac{1}{\G} \left(\sqrt{1+\G} - 1 \right) \simeq 40 \%$ of $v_0$. Similarly, a water diffusing time can be defined $\timeh = l^2/\Dh$, and like inspiratory times and capillary transit times, it varies with $M^{\frac14}$ at rest~\cite{west_general_1997}.}.
This equivalent Fick's law allows to define the water pulmonary diffusing capacity $\Dh$ and the heat pulmonary diffusing capacity $\DH$. These quantities are defined in the same way than the pulmonary diffusing capacity $\Do$ that is related to oxygen transport~\cite{bohr_uber_1909, gehr_design_1981, weibel_pathway_1984}.
The water pulmonary diffusing capacity relates the flow of water leaving the lung with the difference in water partial pressures in air between the trachea and the alveoli~\footnote{Concentrations $C$ and partial pressures $p$ are related by the ideal gas law, $C = p / (R T)$ with $T$ expressed in Kelvin. The concentration difference $\Csat(T_b) - \Cin_0$ is actually proportional to the partial pressure differences corrected with the temperature, i.e 
$$
\Csat(T_b) - \Cin_0 = \frac1{R}\left(\frac{\psat(T_b)}{T_b} - \frac{p^{\rm{in}}_0}{T_0}\right)
$$
However, expressed in Kelvin, the ratio $(T_b - T_0)/T_b$ is generally less than $10 \%$ in typical mammals' environmental conditions. So, we can assume that 
$$
\Csat(T_b) - \Cin_0 \simeq \frac{\psat(T_b) - p^{\rm{in}}_0}{R T_b}
$$
}, 
$
\Vh = \Dh \left( \psat(T_b) - p^{\rm{in}}_0 \right)
$. The water pulmonary diffusing capacity can be written as
\begin{multline}
\Dh = \beta_{DH_2O} \frac{\ve}{\G} \left(\sqrt{1+\G} - 1 \right)\\ 
\simeq \frac{0.82}{\G} \left( \sqrt{1+\G} - 1 \right) \ M^{\frac34} \ \rm{ml/(min.mmHg)}
\end{multline}
with $\ve$ the minute ventilation~\cite{haverkamp_physiologic_2005} and $\beta_{DH_2O} = 1.32 \ 10^{-3} \ \rm{mmHg}^{-1}$ a conversion constant~\footnote{$\beta_{DH_2O} = 133.32 \times \frac{V_M}{R T_b}$, with $V_M$ the molar volume at $38^{\circ}\rm{C}$, $V_M = 25.53 \ 10^{-3} \ \rm{m}^3/\rm{mol}$, $R$ the ideal gas constant, $R = 8.314 \ \rm{J/(K.mol)}$ and $T_b$ the body temperature in Kelvin. We can neglect the variations of $T_b$ in the expression of $\beta_D$, since the range of body temperature in mammals spans on about $10 \ \rm{K}$ only~\cite{white_mammalian_2003}. We choose for $T_b$ the corresponding value in Kelvin for $38^{\circ} \rm{C}$, i.e. $T_b = 311.15 \ \rm{K}$.}. $\Dh$ does depend on $\ve$, at the contrary of $\Do$. In addition, $\Dh \propto M^{\frac34}$ whereas $\Do \propto M^1$~\cite{gehr_design_1981, west_general_1997}. These different scalings highlight that $\Dh$ is driven by the scaling of the air flow, while $\Do$ is driven by the space-filling properties of the respiratory gaz exchange surface~\cite{gehr_design_1981}.
As a consequence, $\Dh / M$ scales with $M^{-\frac14}$ indicating that large mammals at rest tend to lose less water through their lungs, relatively to their mass, than small mammals.

Then, the dissipated power $\P$ can be related to water pressures difference through the definition of a heat pulmonary diffusing capacity $\DH$, allowing to write $\P = \DH \left( \psat(T_b) - p^{\rm{in}}_0 \right)$, where
the expression of the heat pulmonary diffusing capacity is
\begin{multline}
\DH = \L \beta_{D\P} (1+\beta)\frac{\ve}{\G} \left(\sqrt{1+\G} - 1 \right)\\
\simeq 1.69 \ M^{\frac34} \ \rm{J/(min.mmHg)} 
\end{multline}
with $\beta_{D\P} = 5.15 \ 10^{-2} \ \rm{mol/(ml.mmHg)}$~\footnote{$\beta_{D\P} = 133.32 / (R T_b)$, with $R$ the ideal gas constant, $R = 8.314 \ \rm{J/(K.mol)}$ and $T_b = 311.15 \ \rm{K}$ the body temperature.}.
Both pulmonary diffusive capacities might prove to be interesting tools to analyse and compare inter- or intraspecific behaviors either in the biological or in the medical frame.\\ 

\begin{figure}[h!]
\includegraphics[width=7cm]{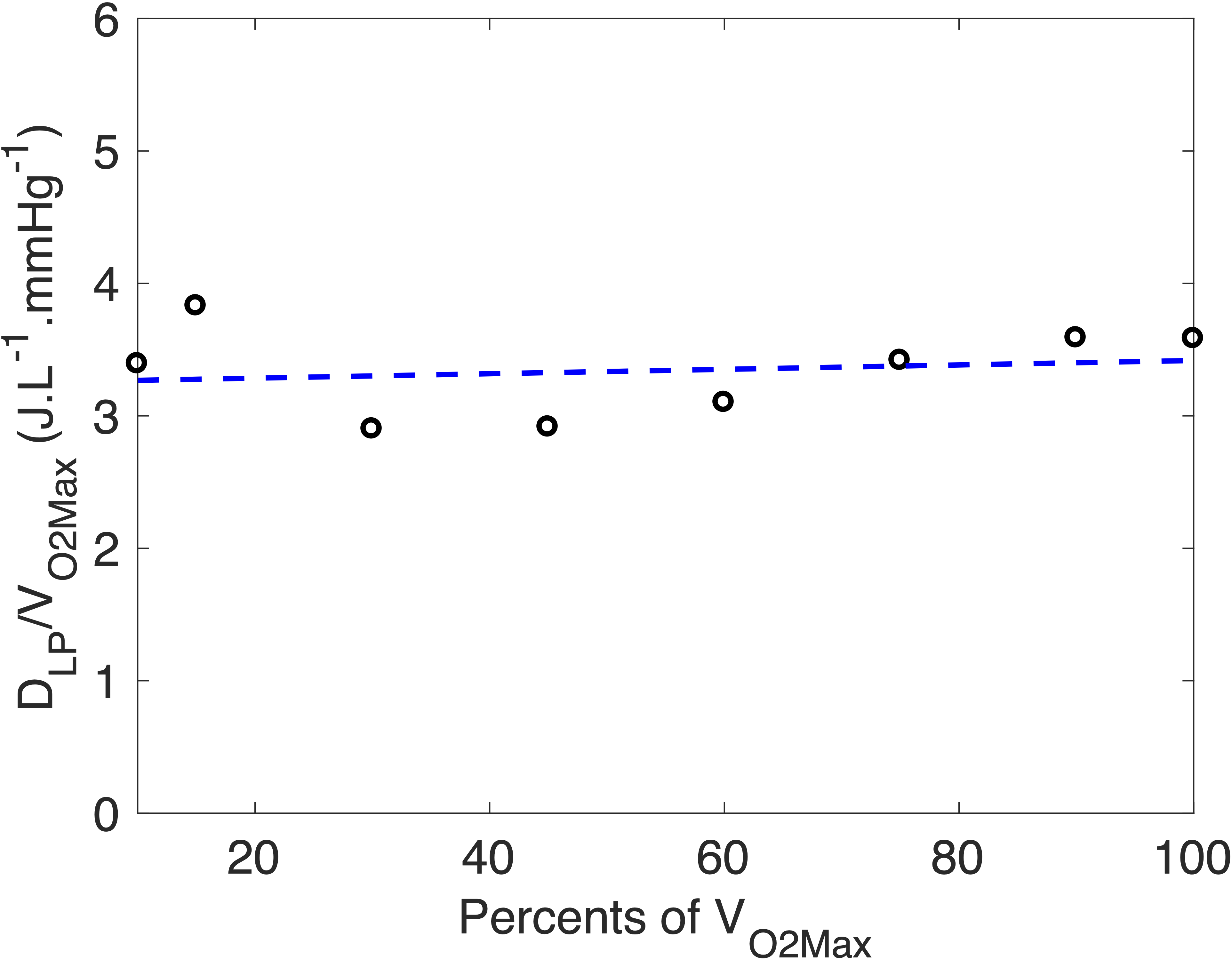}
\caption{Ratio between the heat pulmonary diffusive capacity $\DH$ and $\vo$ in human at different levels of exercise~\cite{haverkamp_physiologic_2005}. The ratio varies only slightly with the exercise intensity, indicating that both quantities are evolving at a similar pace: the proportion of heat dissipated by the lung is predicted to be almost the same whatever the regime.}
\label{Figure4}
\end{figure}

{\bf Exercise in humans.}
The previous results should be reenforced at metabolic regimes higher than rest, typically at field metabolic rate and at maximal metabolic rate~\cite{nagy_field_2005, weibel_exercise-induced_2005}. However, gaps in the literature in term of allometric properties at these regimes prevent from a full extension of our allometric analysis to both field and exercise metabolic rates. Also, our model is probably less accurate at exercise, because more complex flow patterns could affect exchanges at the air--mucosa interface in the proximal bronchi. Nevertheless, an analysis at exercise in human could bring some interesting preliminar insights in an intraspecific framework. The response at maximal metabolic rate goes with an increase of the number $\G$, indicating that $\DH/\ve \propto \frac{1}{\G}(\sqrt{1+\G}-1)$ is decreasing. Hence, the response in term of power diffusing capacity is a trade-off between the decrease of $\frac{1}{\G}(\sqrt{1+\G}-1)$ and the increase of the ventilation $\ve$. We estimated this trade-off in human using data from~\cite{haverkamp_physiologic_2005}. For exercising humans, $\DH$ is increasing with the exercise intensity. Moreover, the ratio $\DH / \vv$ varies only slightly, as shown on Fig.~\ref{Figure4}. Hence, the increase of the heat pulmonary diffusing capacity almost fits the increase in term of energy needs in humans, which itself is related to the body heat production, but probably not in a linear way. At maximal exercise, $\G$ is increased by a factor $3.64$ hence decreasing $\frac{1}{\G}(\sqrt{1+\G}-1)$ from $40 \%$ to $30 \%$. However, the ventilation increases by a factor larger than $14$, inducing a global increase of $\DH$ of about $10$, which is the proportion of increase of $\vv$ from rest to maximal exercise~\cite{haverkamp_physiologic_2005}. This indicates that the heat pulmonary diffusive capacity $\DH$ in human might adjust to the working load of the body, thus always keeping pace with the heat production. 


Our study derives, for the first time, with an analytical, highly tractable and validated model, how a universal number $\G$ drives the heat and water exchanges amongst mammals. The model for the temperature of the mucosa reflected by equation (\ref{Tconj}) allows to get a first theoretical estimation of $\G$ and paves the way to the understanding of the alternative strategies used by mammals to dissipate their excess of heat, depending on their size. Our study supports the idea that the lower mucosa temperature in the proximal bronchi could well not be solely a side effect of respiration, but instead a selected feature of the respiratory system. This looks especially crucial for large mammals, whose lungs plays a relatively more important role in their heat dissipation strategy than small mammals. As shown by our analysis for exercising human, the heat and water pulmonary diffusing capacities that we introduced in this work might constitute powerful tools in the study of inter- and intraspecific properties of the healthy or pathologic lung as a heat and water exchanger. Improving our understanding of how mammals, and more generally endotherms, are able to dissipate heat might bring important insights on mammals' thermoregulation and on the effect of warming environments.
Future works should account for more detailed representation of the air flow patterns in the bronchi and of the temperature distribution inside the bronchi mucosa.

\vspace{1cm}

\begin{acknowledgments}
The authors gratefully acknowledge the financial support of ESA and BELSPO (ESA-ESTEC-PRODEX arrangement 4000109631), Fonds de la Recherche Scientifique - FNRS (Postdoctoral Researcher position of BS) and the French National Research Agency (ANR VirtualChest, ANR-16-CE19-0014 and Idex UCA JEDI, ANR-15-IDEX-01).\end{acknowledgments}


%

\end{document}